\documentclass[%
 reprint,
 amsmath,amssymb,
 aps,prb
]{revtex4-1}

\usepackage{graphicx}
\usepackage{dcolumn}
\usepackage{bm}
\usepackage{hyperref}
\usepackage{float,epstopdf}
\usepackage{blindtext}

\begin{document}

\preprint{APS/123-QED}

\title{Enhancement of induced synchronization by transient uncoupling \\ in coupled simple chaotic systems}
\author{G. Sivaganesh}
\affiliation{%
Department of Physics, Alagappa Chettiar Government College of Engineering $\&$ Technology, Karaikudi, Tamilnadu-630 004, India\\
 }%
\author{A. Arulgnanam}
 \email[Corresponding author: ]{gospelin@gmail.com}
\affiliation{%
Department of Physics, St. John's College, Palayamkottai, Tamilnadu-627 002, India, Affiliated to Manonmaniam Sundaranar University, Abishekapatti, Tirunelveli, Tamilnadu - 627 012, India\\
 }%
\author{A. N. Seethalakshmi}
\affiliation{%
Department of Physics, The M.D.T Hindu College, Tirunelveli, Tamilnadu-627 010, India, Affiliated to Manonmaniam Sundaranar University, Abishekapatti, Tirunelveli, Tamilnadu - 627 012, India\\
 }%

\date{\today}

\begin{abstract}
In this paper, we report the enhanced stability of induced synchronization by transient uncoupling observed in certain unidirectionally coupled second-order chaotic systems. The stability of synchronization observed in the coupled systems subjected to transient uncoupling is analyzed using the {\emph{Master Stability Function}}. The existence of the coupled systems in stable synchronized states over a certain range of the clipping fraction of the driven system is identified. The enhanced stable synchronized states are obtained for fixed values of clipping fraction in certain second-order chaotic systems. The two-parameter bifurcation diagram indicating the parameter regions over which stable synchronization occurs is presented. The negative eigenvalue regions of the driven system enabling induced synchronization is studied for all the systems. The enhancement of synchronization through transient uncoupling observed in coupled second-order non-autonomous chaotic systems is reported in the literature for the first time.
\begin{description}

\item[PACS numbers]
05.45.Gg; 05.45.Xt;
\end{description}
\end{abstract}
                             
\maketitle


\section{INTRODUCTION}

The identical behavior of coupled chaotic systems has been primarily challenged by the sensitive dependence nature of chaotic trajectories on initial conditions. With the identification of chaos synchronization by {\emph{Pecora et al.}} \cite{Pecora1990}, several nonlinear systems have been studied for the dynamical process of chaos synchronization \cite{Ogorzalek1993,Rulkov1995,Rosenblum1996,Boccaletti2002}. The existence of the drive and the driven systems in the synchronized state has been indicated by the negative values of the  {\emph{Master Stability Function}} (MSF) \cite{Pecora1998,Liang2009}. Hence, in order to sustain synchronization in coupled chaotic systems the MSF has been expected to be in the negative valued regions for greater coupling strengths. Recently, induced synchronization has been instigated in unidirectionally coupled chaotic systems by the method of transient uncoupling \cite{Schroder2015}. In the method of transient uncoupling, the coupling strength is activated only over a certain region of phase space of the chaotic attractor of driven system. This method has been applied for a variety of higher dimensional chaotic systems and the mechanism for the emergence of enhanced synchronization has been identified\cite{Schroder2016,Aditya2016,Ghosh2018}. However, the applicability of the method of transient uncoupling for second-order, non-autonomous chaotic systems remains unexplored. In this paper, we present the enhancement of induced synchronization by transient uncoupling observed in certain coupled second-order, non-autonomous chaotic systems.  Enhancing the stability of synchronized states in coupled chaotic systems becomes necessary in understanding the collective dynamics of these systems in networks \cite{Liang2009}. \\
Now, we discuss the method of transient uncoupling in brief. Under the uncoupled state, the isolated systems are described by
\begin{equation}
\bf{\dot x} =   \bf{F(x)}
\label{eqn:1}
\end{equation}
where, {\bf{F(x)}} is the velocity field, {\bf{x}} is a $d$-dimensional vector and $d$ represents the dimension of the system. With the transient uncoupling factor introduced in the coupling parameter, the driven system which is coupled to the drive can be written as
\begin{equation}
\bf{\dot x_2} =   \bf{F(x_2)} + \epsilon \chi_{A} \sum_{j=1}^{N} {\bf{G}_{ij}} {\bf{E}}(x_2),
\label{eqn:2}
\end{equation}
where {\bf{G}} and {\bf{E}} are the matrices representing coupling coefficients and the information of the coupled variables, respectively. The term $\chi_A$ represents the transient uncoupling factor given as
\begin{equation}
\chi_A =
\begin{cases}
1 & \text{if ${\bf{x_2}} \in A$}\\
0 & \text{if ${\bf{x_2}} \notin A$}
\end{cases}
\label{eqn:3}
\end{equation}
where, $A$ is a region of phase space of the driven system such that $A \subseteq \mathbb{R}^d$ in which the driven system is controlled by the drive and the represents the region where the coupling is active. The subset $A$ is obtained by clipping a portion of the phase space of the driven system along the direction of the coupled state variable of the driven system given as
\begin{equation}
A_{\Delta} = \{ {\bf{x}}_2 \in \mathbb{R}^d : |{\bf{x}}_2 - {\bf{x}}_{2}^*| \le \Delta \}
\label{eqn:4}
\end{equation}
where ${\bf{x_2}}^{*}$ is a point considered as the center of the chaotic attractor and it lies along the coordinate axis of the state variable $\bf{x_2}$. The variational equation of Eq. \ref{eqn:2} is given by
\begin{equation}
\dot \xi = [{\bf{I}}_N \otimes D{\bf{F}} + \epsilon \chi_{A} (\bf{G} \otimes  \bf{E})] \xi
\label{eqn:5}
\end{equation}
where $\bf{I}_N$ is an $N \times N$ identity matrix, D{\bf{F}} is the Jacobian of the uncoupled system and $\otimes$ represents the {\emph{inner}} or {\emph{Kronecker}} product. On diagonalization of the matrix {\bf{G}}, Eq. \ref{eqn:5} is written as
\begin{equation}
\dot {\xi_k} = [D{\bf{F}} + \delta \gamma_k \bf{E})] \xi_k,
\label{eqn:6}
\end{equation}
where $\delta = \epsilon \chi_{A}$ and $\gamma_k$ are the eigenvalues of {\bf{G}} with {\emph{k}} = 0 or 1. In general, the quantity $\delta \gamma_k$ are generally complex numbers which can be written in the form $\delta \gamma_k = \alpha + i \beta$. Hence, the general dynamical system is
\begin{equation}
\dot {\xi_k} = [D{\bf{F}} + (\alpha + i \beta) \bf{E})] \xi_k,
\label{eqn:7}
\end{equation}
The largest transverse lyapunov exponent $\lambda^{\perp}_{max}$ of the generic variational equation given by Eq. \ref{eqn:7} depending on $\alpha$ and $\beta$ values, is the {\emph{master stability function}} (MSF) \cite{Pecora1998,Pecora1997}. Under the coupling of the $x$-variables of the system, the matrices $\bf G$ and $\bf E$ are given as
\begin{equation*}
\bf{G} =
\begin{pmatrix}
-1 &&& 1 \\
 0 &&&  0 \\
\end{pmatrix},~
\bf{E} =
\begin{pmatrix}
1 &&& 0 &&& 0\\
0 &&& 0 &&& 0 \\
0 &&& 0 &&& 0\\
\end{pmatrix}. 
\end{equation*} 
The following are discussed in this paper. In Section \ref{sec:2}, the enhancement of stable synchronization observed in coupled {\emph{Morse}} oscillator systems subjected to transient uncoupling is presented. Section \ref{sec:3} deals with the studies on coupled series LCR circuit systems with two different types of nonlinear elements and in Section \ref{sec:4} coupled parallel LCR circuit systems are discussed. In Section \ref{sec:5}, the enhancement of synchronization observed in  coupled quasiperiodically forced series LCR circuit systems exhibiting strange non-chaotic attractors (SNA) in their dynamics is presented. 

\section{Morse Oscillator}
\label{sec:2}
\begin{figure}
\begin{center}
\includegraphics[scale=0.66]{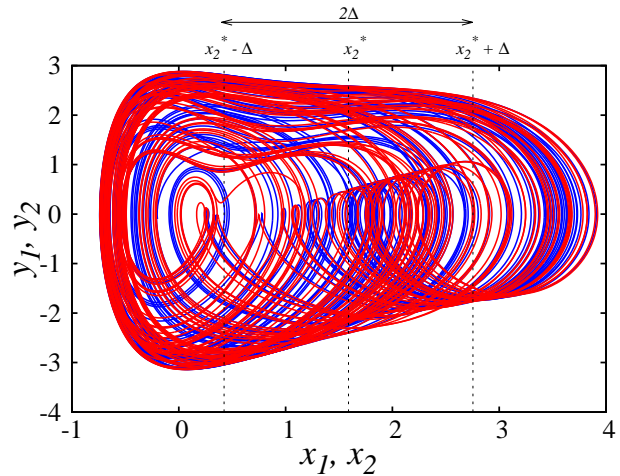}
\caption{(Color Online) Chaotic attractors of the drive (blue) and driven systems (red) of the coupled {\emph{Morse}} oscillator system in the $x_{1}-y_{1}$ and $x_{2}-y_{2}$ planes under the uncoupled state $(\epsilon = 0)$ with clipping of phase space of the driven system through transient uncoupling.}
\label{fig:1}
\end{center}
\end{figure}

The classical {\emph{Morse}} oscillator system has been used to describe the dynamics of diatomic molecular vibrations. The dynamical equations of the unidirectionally coupled {\emph{Morse}} oscillator system \cite{George1986,Lakshmanan2003} under transient uncoupling is written as
\begin{subequations}
\begin{eqnarray}
\dot x_1  &=&  y_1, \\ 
\dot y_1  &=&  -\alpha y_1 - \beta e^{-x_1} (1-e^{-x_1}) + f_1~cos(z_1),\\ 
\dot z_1  &=&  \omega_1,\\
\dot x_2  &=&  y_2+ \epsilon \chi_A (x_1 - x_2), \\ 
\dot y_2  &=&  -\alpha y_2 - \beta e^{-x_2} (1-e^{-x_2}) + f_2~cos(z_2),\\ 
\dot z_2  &=&  \omega_2,
\end{eqnarray}
\end{subequations}
where $x_1,y_1,z_1$ and $x_2,y_2,z_2$ represents the normalized state variables of the drive and driven systems, respectively. The drive and the driven systems are coupled through the $x$-variables and the parameters of the system are fixed as $\alpha = 0.8,~\beta = 8,~\omega_{1,2}=2$ and $f_{1,2}=0.35$. The {\emph{Morse}} oscillator system exhibits a period-doubling route to chaos and the chaotic attractors of the drive (blue) and the driven (red) systems obtained for the amplitude of the external force $f_{1,2}=0.35$ under the unsynchronized state $(\epsilon=0)$ is as shown in Fig. \ref{fig:1}.
\begin{figure}
\begin{center}
\includegraphics[scale=0.33]{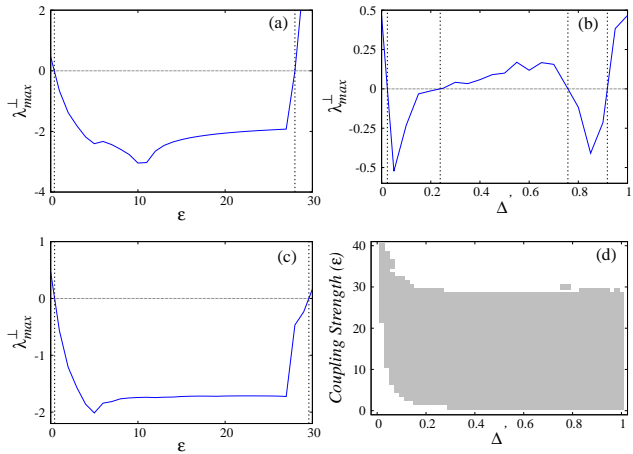}
\caption{(Color Online) Stability of synchronization observed in coupled {\emph{Morse}} oscillator systems. (a) MSF obtained without transient uncoupling exists in negative valued regions in the range $0.415 \le \epsilon \le 28$; (b) MSF as a function of $\Delta^{'}$ for $\epsilon = 29.5$ indicating stable synchronized states in the lower and higher regions of clipping fractions $0.0235 \le \Delta^{'} \le 0.2383$ and $0.7576 \le \Delta^{'} \le 0.9179$, respectively; (c) Variation of MSF with $\epsilon$ under transient uncoupling obtained with ${x_2}^{*} = 1.59$ and $\Delta^{'} = 0.85$ indicating enhanced stable synchronized states in the range of coupling strengths $0.45 \le \epsilon \le 29.6$; (d) Two-parameter bifurcation diagram in the $\Delta^{'}-\epsilon$ plane indicating the parameter regions for stable synchronized states (gray color).}
\label{fig:2}
\end{center}
\end{figure}
\begin{figure}
\begin{center}
\includegraphics[scale=0.66]{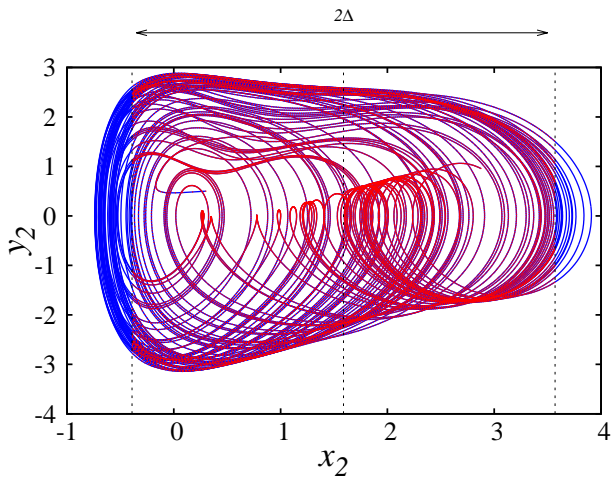}
\caption{(Color Online) Chaotic attractors of the driven {\emph{Morse}} oscillator system (blue) in the $x_2-y_2$ phase space with regions of $x_2$ (red) for which the real parts of the eigenvalues of the driven system is negative under a finite clipping fraction for the parameters ${x_2}^{*} = 1.59$ and $\Delta^{'} = 0.85$.}
\label{fig:3}
\end{center}
\end{figure}
The synchronization dynamics of the coupled system under transient uncoupling is shown in Fig. \ref{fig:2}. In the absence of transient uncoupling the variation of the MSF as a function of the coupling parameter shows stable synchronized states in the range $0.415 \le \epsilon \le 28$. Hence, the coupled system becomes unsynchronized for coupling strengths $\epsilon > 28$ under normal coupling. Now we introduce the transient uncoupling factor in the system equations and study its synchronization stability. The stability of the synchronized states can be analyzed for different regions of clipping of the phase space of the driven system expressed by the clipping fraction $\Delta^{'} = 2\Delta/\Omega$ where, $\Omega$ is the width of the attractor along the clipping coordinate axis (x-axis). The point ${x_2}^*$ has been considered as the center of the chaotic attractor and the phase space is clipped to a width of $\Delta$ along the coordinate axis ($x_2$-axis) on either side of ${x_2}^*$, resulting in a clipping width of $2\Delta$. The variation of the MSF of the coupled system  under transient uncoupling as a function of the clipping fraction with the coupling strength fixed at $\epsilon = 29.5$ is shown in Fig. \ref{fig:2}(b). For $\Delta=0$, the two systems are uncoupled and hence does not synchronize and for $\Delta = \Omega/2$, the coupling between the systems become normal and the original unsynchronized behavior of the systems under normal coupling is obtained. From Fig. \ref{fig:2}(b) it is observed that the coupled systems exist in stable synchronized states in the range of the clipping fractions $0.0235 \le \Delta^{'} \le 0.2383$ and $0.7576 \le \Delta^{'} \le 0.9179$, respectively. Figure \ref{fig:2}(c) showing the variation of the MSF with the coupling strength indicates a broader stable synchronized state in the range $0.45 \le \epsilon \le 29.6$ for a clipping fraction of $\Delta^{'}=0.85$. The introduction of transient uncoupling induces synchronization at the values of higher coupling strength for which synchronization is not been observed under normal coupling. The two-parameter bifurcation diagram obtained in the $\Delta^{'}-\epsilon$ plane showing the parameter regions (gray) for the existence of the coupled system in the synchronized state is as shown in Fig. \ref{fig:2}(d).\\

An analysis on the nature of the eigenvalues of the driven system within the clipped region provides an understanding on the mechanism of induced synchronization through transient uncoupling. The Jacobian of the coupled {\emph{Morse}} oscillator system subjected to transient uncoupling can be written as

\setcounter{MaxMatrixCols}{12}
\begin{equation}
J =
\begin{pmatrix}
0 && 1 && 0 && 0 \\
\beta e^{-x_1} (1-e^{-x_1}) && -\alpha && 0 && 0 \\
\epsilon \chi_{A} && 0 && -\epsilon \chi_{A} && 1 \\
0 && 0 && \beta e^{-x_2} (1-e^{-x_2}) && -\alpha \\
\end{pmatrix}. 
\label{eqn:9}
\end{equation} 

The synchronous of the driven system is observed only when the trajectories of the driven system traces that of the drive. Hence the eigenvalues corresponding to the driven system obtained from Eq. \ref{eqn:9} must enable the trajectories to converge towards the drive system. Figure \ref{fig:3} shows the chaotic attractor (blue) of the driven system in the $(x_2-y_2)$ phase plane along with the values of $x_2$ (red) within the clipping fraction $\Delta^{'}=0.85$ for which the eigenvalues of the driven system have negative real parts and hence leading to the convergence of the trajectories. These negative eigenvalue regions of the driven system within the clipping fraction enables the synchronization of the coupled systems.

\section{Series LCR circuits}
\label{sec:3}
In this section, we present the enhanced stability of induced synchronization observed in a class of coupled sinusoidally forced series LCR circuit systems with three-segmented piecewise-linear elements. The {\emph{Chua's diode}} and the {\emph{simplified nonlinear element}} are the nonlinear elements considered for the present study. The normalized state equations of the unidirectionally coupled chaotic systems subjected to transient uncoupling can be written as 
\begin{subequations}
\begin{eqnarray}
\dot x_1  &=&  y_1 - h(x_1), \\ 
\dot y_1  &=&  -\beta y_1 - \nu \beta y_1 - \beta x_1 + f_1 sin(z_1) ,\\ 
\dot z_1  &=&  \omega_1,\\
\dot x_2  &=&  y_2 - h(x_2)+ \epsilon \chi_A (x_1 - x_2), \\ 
\dot y_2  &=&  -\beta y_2 - \nu \beta y_2 - \beta x_2 + f_2 sin(z_2) ,\\ 
\dot z_2  &=&  \omega_2,
\end{eqnarray}
\label{eqn:10}
\end{subequations}
where $x_1,y_1,z_1$ and $x_2,y_2,z_2$ represents the normalized state variables of the drive and driven systems, respectively. The terms $h(x_{1}), h(x_2)$ represent the three-segmented piecewise-linear function of the drive and the driven systems given as 
\begin{equation}
h(x_{1,2}) =
\begin{cases}
bx_{1,2}+(a-b) & \text{if $x_{1,2} \ge 1$}\\
ax_{1,2} & \text{if $|x_{1,2}| \le1$}\\
bx_{1,2}-(a-b) & \text{if $x_{1,2} \le -1$}
\end{cases}
\label{eqn:11}
\end{equation}
The driven system is coupled to the drive through the $x$-variable by the factor $\epsilon \chi_A$. When the piecewise-linear function $h(x)$ represents the characteristics of the {\emph{Chua's diode}} the circuit is termed as the {\emph{Murali-Lakshmanan-Chua circuit}} (MLC) circuit and when it represents the {\emph{simplified nonlinear element}} it is termed as the forced series LCR circuit with a {\emph{simplified nonlinear element}}. The parameters for the MLC and the circuit with a {\emph{simplified nonlinear element}} take the values $a=-1.02,~b=-0.55,~\beta=1,~\nu=0.015,~\omega_{1,2}=0.72$ and $a=-1.148,~b=5.125,~\beta=0.9865,~\nu=0,~\omega_{1,2}=0.7084$, respectively.

\subsection{Murali-Lakshmanan-Chua circuit}
\label{sec:3.1}

\begin{figure}
\begin{center}
\includegraphics[scale=0.33]{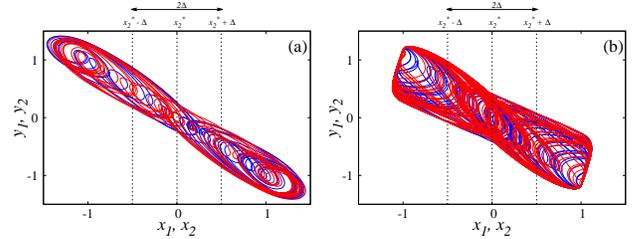}
\caption{(Color Online) Chaotic attractors of the drive (blue) and driven systems (red) of the series LCR circuit systems with a three-segmented piecewise-linear element in the $x_{1}-y_{1}$ and $x_{2}-y_{2}$ planes under the uncoupled state $(\epsilon = 0)$ with clipping of phase space of the driven system through transient uncoupling.  (a) The {\emph{Murali-Lakshmanan-Chua}} circuit system and (b) the forced series LCR circuit system with a {\emph{simplified nonlinear element}}.}
\label{fig:4}
\end{center}
\end{figure}

\begin{figure}
\begin{center}
\includegraphics[scale=0.33]{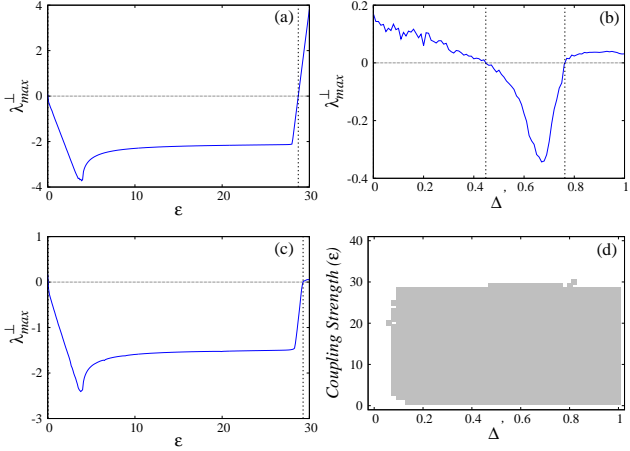}
\caption{(Color Online) Stability of synchronization observed in coupled {\emph{Murali-Lakshmanan-Chua}} circuits. (a) MSF obtained without transient uncoupling exists in negative valued regions in the range $0.0256 \le \epsilon \le 28.725$; (b) MSF as a function of $\Delta^{'}$ for $\epsilon = 29$ indicating stable synchronized states in the range of clipping fractions $0.4478 \le \Delta^{'} \le 0.7624$; (c) Variation of MSF with $\epsilon$ under transient uncoupling obtained with ${x_2}^{*} = 0.0005$ and $\Delta^{'} = 0.67$ indicating enhanced stable synchronized states in the range of coupling strengths $0.0528 \le \epsilon \le 29.277$; (d) Two-parameter bifurcation diagram in the $\Delta^{'}-\epsilon$ plane indicating the parameter regions for stable synchronized states (gray color).}
\label{fig:5}
\end{center}
\end{figure}

The {\emph{Murali-Lakshmanan-Chua circuit}} circuit is the first second-order, non-autonomous circuit system found to exhibit chaos in its dynamics \cite{Murali1994a}. The chaotic and synchronization dynamics of the circuit has been completely studied experimentally, numerically and analytically \cite{Murali1995,Sivaganesh2015,Sivaganesh2018,Sivaganesh2019}. This system exhibits a double-band chaotic attractor at the amplitude of the external force $f_{1,2} = 0.14$. The chaotic attractors of the drive (blue) and the driven (red) MLC circuit systems under the unsynchronized state is as shown in Fig. \ref{fig:4}(a). The synchronization dynamics of the coupled system under transient uncoupling is shown in Fig. \ref{fig:5}. Under normal coupling, the variation of the MSF with the coupling strength indicates stable synchronized states in the range $0.0256 \le \epsilon \le 28.725$ as shown in Fig. \ref{fig:5}(a). Hence, the coupled system becomes unsynchronized for coupling strengths $\epsilon > 28.725$ under normal coupling. The variation of the MSF of the coupled system under transient uncoupling as a function of the clipping fraction with the coupling strength fixed at $\epsilon = 29$ is shown in Fig. \ref{fig:5}(b). From Fig. \ref{fig:5}(b) it is observed that the coupled system exists in stable synchronized states in the range of the clipping fractions $0.4478 \le \Delta^{'} \le 0.7624$. Figure \ref{fig:5}(c) showing the variation of the MSF with the coupling strength indicates stable synchronized state in the range $0.0528 \le \epsilon \le 29.277$ for a clipping fraction of $\Delta^{'}=0.67$. The two-parameter bifurcation diagram obtained in the $\Delta^{'}-\epsilon$ plane showing the parameter regions (gray) for the existence of the coupled system in the synchronized state is shown in Fig. \ref{fig:5}(d). 
\begin{figure}
\begin{center}
\includegraphics[scale=0.33]{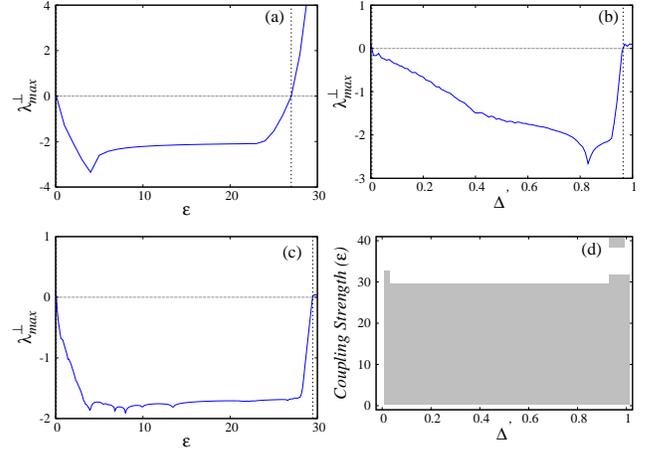}
\caption{(Color Online) Stability of synchronization observed in coupled forced series LCR circuit systems with {\emph{simplified nonlinear element}}. (a) MSF obtained without transient uncoupling exists in negative valued regions in the range $0.074 \le \epsilon \le 27.006$; (b) MSF as a function of $\Delta^{'}$ for $\epsilon = 29$ indicating the stable synchronized states in the range of clipping fractions $0.004 \le \Delta^{'} \le 0.964$; (c) Variation of MSF with $\epsilon$ under transient uncoupling obtained with ${x_2}^{*} = 0.0004$ and $\Delta^{'} = 0.5$ indicating enhanced stable synchronized states in the range of coupling strengths $0.067 \le \epsilon \le 29.467$; (d) Two-parameter bifurcation diagram in the $\Delta^{'}-\epsilon$ plane indicating the parameter regions for stable synchronized states (gray color).}
\label{fig:6}
\end{center}
\end{figure}
\begin{figure}
\begin{center}
\includegraphics[scale=0.33]{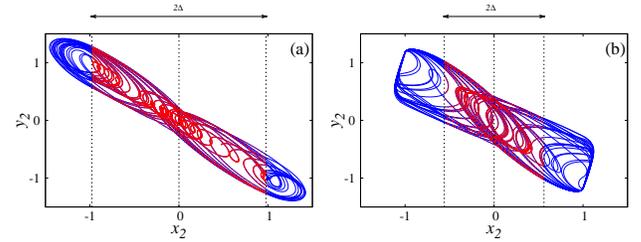}
\caption{(Color Online) (Color Online) Chaotic attractors of the response system (blue) in the $x_2-y_2$ phase space with regions of $x_2$ (red) for which the real parts of the eigenvalues of the driven system is negative under a finite clipping fraction. (a)  {\emph{Murali-Lakshmanan-Chua}} obtained for the parameters ${x_2}^{*} = 0.0005$ and $\Delta^{'} = 0.67$; (b) Forced series LCR circuit with {\emph{simplified nonlinear element}} obtained for the parameters ${x_2}^{*} = 0.0004$ and $\Delta^{'} = 0.5$.}
\label{fig:7}
\end{center}
\end{figure}
\subsection{Forced series LCR circuit with a simplified nonlinear element}
\label{sec:3.2}

The forced series LCR circuit with a {\emph{simplified nonlinear element}}  introduced by {\emph{Arulgnanam et al.}} \cite{Arulgnanam2009} is a second-order, non-autonomous circuit that exhibits chaos in its dynamics. The {\emph{simplified nonlinear element}} is an active circuit element that has been constructed with a least number of circuit elements. The fractal dimension of the chaotic attractors produced by the circuits with this nonlinear element has a greater value as compared to the {\emph{Chua's diode}}. This circuit exhibits a prominent chaotic attractor at the value of the amplitude of the force $f_{1,2}=0.375$. The chaotic attractors of the drive (blue) and the driven (red) systems represented by Eq. \ref{eqn:10} under the unsynchronized state is as shown in Fig. \ref{fig:4}(b). \\
The synchronization dynamics of the coupled system under transient uncoupling is shown in Fig. \ref{fig:6}. Under normal coupling, the variation of the MSF with the coupling strength indicates stable synchronized states in the range $0.074 \le \epsilon \le 27.006$ as shown in Fig. \ref{fig:6}(a). The variation of the MSF of the coupled system under transient uncoupling as a function of the clipping fraction with the coupling strength fixed at $\epsilon = 29$ is shown in Fig. \ref{fig:6}(b). From Fig. \ref{fig:6}(b) it is observed that the coupled system exists in stable synchronized states in the range of the clipping fractions $0.004 \le \Delta^{'} \le 0.964$. Figure \ref{fig:6}(c) showing the variation of the MSF with the coupling strength indicates stable synchronized states in the range $0.067 \le \epsilon \le 29.467$ for a clipping fraction of $\Delta^{'}=0.5$. The two-parameter bifurcation diagram obtained in the $\Delta^{'}-\epsilon$ plane showing the parameter regions (gray) for the existence of the coupled system in the synchronized state is shown in Fig. \ref{fig:6}(d). \\
The Jacobian of the normalized state equations of the series LCR circuit systems with any three-segmented piecewise-linear element represented by Eq. \ref{eqn:10} can be written as
\setcounter{MaxMatrixCols}{12}
\begin{equation}
J =
\begin{pmatrix}
-h^{'}(x_1) && 1 && 0 && 0 \\
-\beta && -\sigma && 0 && 0 \\
\epsilon \chi_{A} && 0 && -\epsilon \chi_{A}-h^{'}(x_2) && 1 \\
0 && 0 && -\beta && -\sigma \\
\end{pmatrix}. 
\label{eqn:12}
\end{equation} 
The eigenvalues of the driven systems obtained using Eq. \ref{eqn:12} can be used to analyze the mechanism of induced synchronization by transient uncoupling. Figure \ref{fig:7}(a) and \ref{fig:7}(b) shows the chaotic attractors (blue) of the driven MLC and the {\emph{simplified nonlinear element}} systems in the $(x_2-y_2)$ phase plane along with the values of $x_2$ (red) for the parameters $\epsilon = 29,~\Delta^{'}=0.67$ and $\epsilon=29,~\Delta^{'}=0.5$, respectively.  The eigenvalues corresponding to each of the piecewise-linear regions within the clipping fraction are either real or complex conjugates with negative real parts. Hence, the trajectories of the driven system converges towards the trajectories of the drive at greater values coupling strengths using transient uncoupling for which synchronization cannot be achieved under normal coupling.

\section{Parallel LCR circuits}
\label{sec:4}

In this section, we present the enhanced stability of induced synchronization observed in coupled sinusoidally forced parallel LCR circuit systems with piecewise-linear elements. The {\emph{Chua's diode}} and the {\emph{simplified nonlinear element}} are the nonlinear elements considered for the present study. The normalized state equations of the $x$-coupled chaotic systems subjected to transient uncoupling can be written as 
\begin{subequations}
\begin{eqnarray}
\dot x_1  &=&   f_1 sin(z_1) - x_1 - y_1 - h(x_1), \\ 
\dot y_1  &=&  \beta x_1, \\ 
\dot z_1  &=&  \omega_1,\\
\dot x_2  &=&  f_2 sin(z_2) - x_2 - y_2 - h(x_2) + \epsilon \chi_A (x_1 - x_2), \\ 
\dot y_2  &=&  \beta x_2, \\ 
\dot z_2  &=&  \omega_2,
\end{eqnarray}
\label{eqn:13}
\end{subequations}
where $x_1,y_1,z_1$ and $x_2,y_2,z_2$ represents the normalized state variables of the drive and driven systems, respectively. The terms $h(x_{1}), h(x_2)$ representing the three-segmented piecewise-linear function is as given in Eq. \ref{eqn:11}. The driven system is coupled to the drive through the $x$-variable by the factor $\epsilon \chi_A$. When the piecewise-linear function $h(x)$ represents the characteristics of the {\emph{Chua's diode}} the circuit is termed the {\emph{Variant of Murali-Lakshmanan-Chua}} (MLCV) circuit and when it represents the {\emph{simplified nonlinear element}} it is termed the forced parallel LCR circuit with a {\emph{simplified nonlinear element}}. The parameters for the MLCV and the circuit with a {\emph{simplified nonlinear element}} take the values $a=-1.121,~b=-0.6047,~\beta=0.05,~\omega_{1,2}=0.105$ and $a=-1.148,~b=5.125,~\beta=0.2592,~\omega_{1,2}=0.2402$, respectively.

\subsection{Variant of Murali-Lakshmanan-Chua circuit}
\label{sec:4.1}

\begin{figure}
\begin{center}
\includegraphics[scale=0.33]{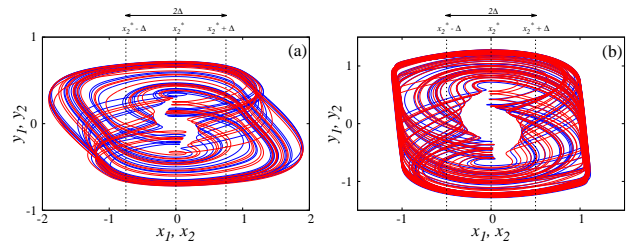}
\caption{(Color Online) Chaotic attractors of the drive (blue) and driven systems (red) of the parallel LCR circuit systems with three-segmented piecewise-linear elements in the $x_{1}-y_{1}$ and $x_{2}-y_{2}$ planes under the uncoupled state $(\epsilon = 0)$ with clipping of phase space of the driven system through transient uncoupling.  (a) The variant of {\emph{Murali-Lakshmanan-Chua}} circuit system and (b) the forced parallel LCR circuit system with a {\emph{simplified nonlinear element}}.}
\label{fig:8}
\end{center}
\end{figure}

\begin{figure}
\begin{center}
\includegraphics[scale=0.33]{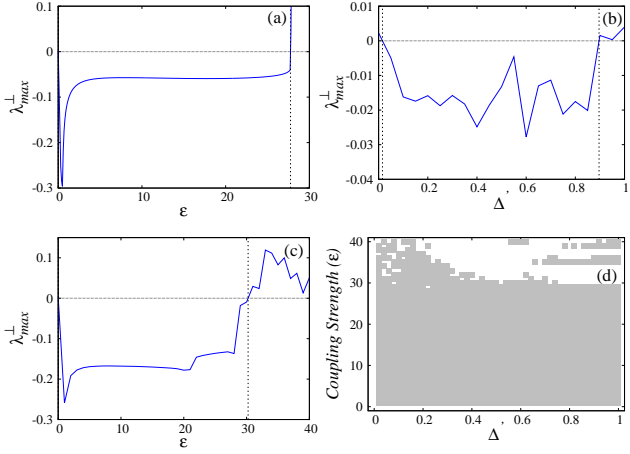}
\caption{(Color Online) Stability of synchronization observed in coupled {\emph{Variant of Murali-Lakshmanan-Chua}} circuits. (a) MSF obtained without transient uncoupling exists in negative valued regions in the range $0.0028 \le \epsilon \le 27.746$; (b) MSF as a function of $\Delta^{'}$ for $\epsilon = 29.25$ indicating the stable synchronized states in the range of clipping fractions $0.0158 \le \Delta^{'} \le 0.8964$; (c) Variation of MSF with $\epsilon$ under transient uncoupling obtained with ${x_2}^{*} = -0.0025$ and $\Delta^{'} = 0.6$ indicating enhanced stable synchronized states in the range of coupling strengths $0.0087 \le \epsilon \le 30.2414$; (d) Two-parameter bifurcation diagram in the $\Delta^{'}-\epsilon$ plane indicating the parameter regions for stable synchronized states (gray color).}
\label{fig:9}
\end{center}
\end{figure}

The forced parallel LCR circuit with the {\emph{Chua's diode}} as the nonlinear element introduced by {\emph{Thamilmaran et al.}} \cite{Thamilmaran2000} exhibits a rich variety of bifurcations and chaos in its dynamics\cite{Thamilmaran2001}. The chaotic and the synchronization dynamics of the circuit has been extensively studied experimentally, numerically and analytically \cite{Sivaganesh2018,Sivaganesh2019,Sivaganesh2017}. This system exhibits a torus-breakdown route to chaos and a prominent chaotic attractor is observed at the amplitude of the external force $f_{1,2} = 0.375$. The chaotic attractors of the drive (blue) and the driven (red) MLCV circuit systems under the unsynchronized state is as shown in Fig. \ref{fig:8}(a). The synchronization dynamics of the coupled system under transient uncoupling is shown in Fig. \ref{fig:9}. Under normal coupling, the variation of the MSF with the coupling strength indicates stable synchronized states in the range $0.0028 \le \epsilon \le 27.746$ as shown in Fig. \ref{fig:9}(a). The variation of the MSF of the coupled system under transient uncoupling as a function of the clipping fraction with the coupling strength fixed at $\epsilon = 29.25$ as shown in Fig. \ref{fig:9}(b) indicates stable synchronized states in the range of the clipping fractions $0.0158 \le \Delta^{'} \le 0.8964$. Figure \ref{fig:9}(c) shows the variation of the MSF with the coupling strength indicating enhanced stable synchronized states in the range $0.0087 \le \epsilon \le 30.2414$ for a clipping fraction of $\Delta^{'}=0.6$. The two-parameter bifurcation diagram obtained in the $\Delta^{'}-\epsilon$ plane showing the parameter regions (gray) for the existence of the coupled system in the synchronized state is as shown in Fig. \ref{fig:9}(d).
\begin{figure}
\begin{center}
\includegraphics[scale=0.33]{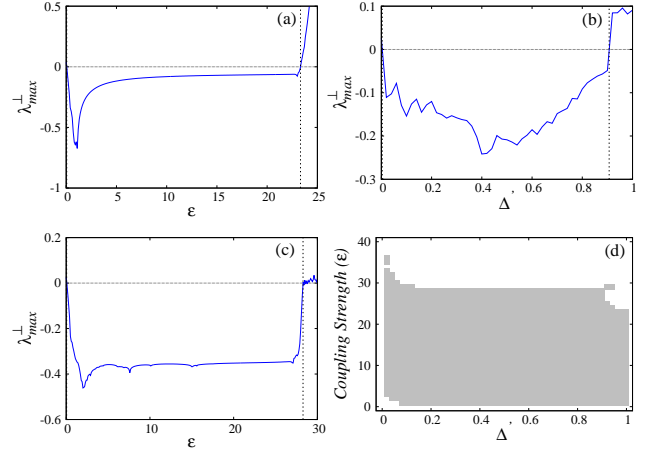}
\caption{(Color Online) Stability of synchronization observed in coupled forced parallel LCR circuits with a {\emph{simplified nonlinear element}}. (a) MSF obtained without transient uncoupling exists in negative valued regions in the range $0.051 \le \epsilon \le 23.32$; (b) MSF as a function of $\Delta^{'}$ for $\epsilon = 28$ indicating stable synchronized states in the range of clipping fractions $0.0037 \le \Delta^{'} \le 0.907$; (c) Variation of MSF with $\epsilon$ under transient uncoupling obtained with ${x_2}^{*} = 0.0006$ and $\Delta^{'} = 0.6$ indicating enhanced stable synchronized states in the range of coupling strengths $0.053 \le \epsilon \le 28.28$; (d) Two-parameter bifurcation diagram in the $\Delta^{'}-\epsilon$ plane indicating the parameter regions for stable synchronized states (gray color).}
\label{fig:10}
\end{center}
\end{figure}
\begin{figure}
\begin{center}
\includegraphics[scale=0.33]{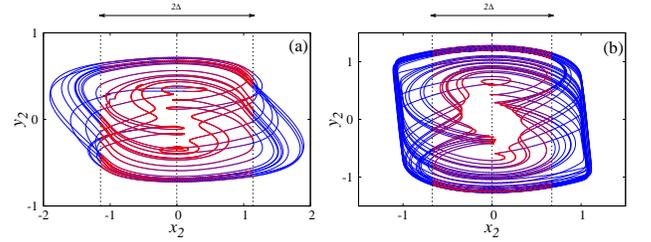}
\caption{(Color Online) (Color Online) Chaotic attractors of the driven system (blue) in the $x_2-y_2$ phase space with regions of $x_2$ (red) for which the real parts of the eigenvalues of the driven system is negative under a finite clipping fraction. (a)  {\emph{Variant of Murali-Lakshmanan-Chua}} circuit obtained for the parameters ${x_2}^{*} = -0.0025$ and $\Delta^{'} = 0.6$; (b) Forced parallel LCR circuit with a {\emph{simplified nonlinear element}} obtained for the parameters ${x_2}^{*} = 0.0006$ and $\Delta^{'} = 0.6$.}
\label{fig:11}
\end{center}
\end{figure}
\subsection{Forced parallel LCR circuit with a simplified nonlinear element}
\label{sec:3.2}

The sinusoidally forced parallel LCR circuit with a {\emph{simplified nonlinear element}} exhibiting chaotic dynamics was introduced by {\emph{Arulgnanam et al.}} \cite{Arulgnanam2015}. This circuit exhibits a prominent chaotic attractor at the amplitude of the force $f_{1,2}=0.695$. The chaotic and the synchronization dynamics of the circuit has been studied experimentally, numerically and analytically \cite{Sivaganesh2018,Sivaganesh2017}. The chaotic attractors of the drive (blue) and the driven (red) systems with {\emph{simplified nonlinear elements}} represented by Eq. \ref{eqn:13} under the unsynchronized state is as shown in Fig. \ref{fig:8}(b). The synchronization dynamics of the coupled system under transient uncoupling is shown in Fig. \ref{fig:10}. Under normal coupling, the variation of the MSF with the coupling strength indicates stable synchronized states in the range $0.051 \le \epsilon \le 23.32$ as shown in Fig. \ref{fig:10}(a). The variation of the MSF of the coupled system under transient uncoupling as a function of the clipping fraction with the coupling strength fixed at $\epsilon = 28$ as shown in Fig. \ref{fig:10}(b) indicates stable synchronized states in the range of the clipping fractions $0.0037 \le \Delta^{'} \le 0.907$. Figure \ref{fig:10}(c) showing the variation of the MSF with the coupling strength indicates stable synchronized state in the range $0.053 \le \epsilon \le 28.28$ for a clipping fraction of $\Delta^{'}=0.6$. The two-parameter bifurcation diagram obtained in the $\Delta^{'}-\epsilon$ plane showing the parameter regions (gray) for the existence of the coupled system in the synchronized state is shown in Fig. \ref{fig:10}(d). \\

The Jacobian of the normalized state equations of the forced parallel LCR circuit systems with a three-segmented piecewise-linear element represented by Eq. \ref{eqn:13} can be written as
\setcounter{MaxMatrixCols}{12}
\begin{equation}
J =
\begin{pmatrix}
-1 - h^{'}(x_1) && -1 && 0 && 0  \\
\beta && 0 && 0 && 0 \\
\epsilon \chi_{A} && 0 && -\epsilon \chi_{A} - 1 - h^{'}(x_2) && -1 \\
0 && 0 && \beta && 0 \\
\end{pmatrix}. 
\label{eqn:14}
\end{equation} 
The eigenvalues of the driven systems obtained from Eq. \ref{eqn:14} can be used to analyze the mechanism of induced synchronization by transient uncoupling. Figure \ref{fig:11}(a) and \ref{fig:11}(b) shows the chaotic attractors (blue) of the driven MLCV and the {\emph{simplified nonlinear element}} systems in the $(x_2-y_2)$ phase plane along with the values of $x_2$ (red) corresponding to negative eigenvalues for the parameters $\epsilon = 28,~\Delta^{'}=0.6$.  The eigenvalues of the driven system corresponding to each of the piecewise-linear regions within the clipping fraction are either real or complex conjugates with negative real parts and hence the trajectories of the driven system converges towards the trajectories of the drive inducing synchronization.

\section{Quasiperiodically forced series LCR circuit}
\label{sec:5}

\begin{figure}
\begin{center}
\includegraphics[scale=0.66]{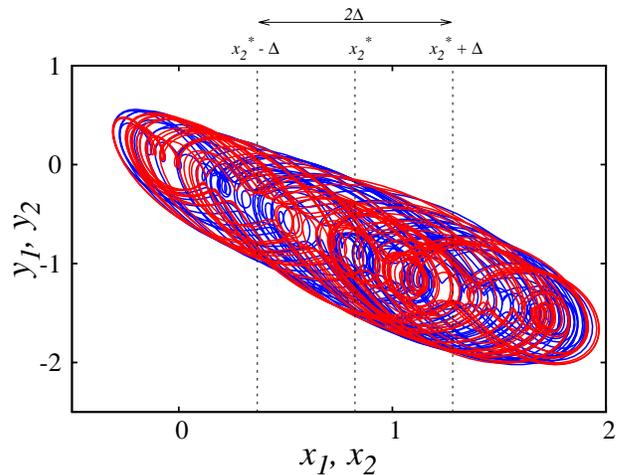}
\caption{(Color Online) Chaotic attractors of the drive (blue) and driven systems (red) of the quasiperiodically forced series LCR circuits with the {\emph{Chua's diode}} as the nonlinear element in the $x_{1}-y_{1}$ and $x_{2}-y_{2}$ planes under the uncoupled state $(\epsilon = 0)$ with clipping of phase space of the driven system through transient uncoupling.}
\label{fig:12}
\end{center}
\end{figure}

The quasiperiodically forced series LCR circuit exhibiting {\emph{strange non-chaotic attractor}} (SNA) in its dynamics was introduced by {\emph{Venkatesan et al.}}\cite{Venkatesan1999}. The synchronization of the SNA dynamics observed in this circuit has been studied both numerically and analytically \cite{Sivaganesh2016,Sivaganesh2016a}. The normalized state equations of the $x$-coupled quasiperiodically forced series LCR circuits with the {\emph{Chua's diode}} as the nonlinear element subjected to transient uncoupling can be written as
\begin{subequations}
\begin{eqnarray}
\dot x_1  &=&  y_1 - h(x_1), \\ 
\dot y_1  &=&  -\beta y_1 - \nu \beta y_1 - \beta x_1 + f_1 sin \theta_1 + f_2 sin \phi_1,\\ 
\dot \theta_1  &=&  \omega_1, \\
\dot \phi_1 &=& \omega_2, \\
\dot x_2  &=&  y_2 - h(x_2)+ \epsilon \chi_A (x_1 - x_2), \\ 
\dot y_2  &=&  -\beta y_2 - \nu \beta y_2 - \beta x_2 + f_3 sin \theta_2 + f_4 sin \phi_2  ,\\ 
\dot \theta_2  &=&  \omega_3, \\
\dot \phi_2 &=& \omega_4, 
\end{eqnarray}
\label{eqn:15}
\end{subequations}
where $x_1,y_1,z_1$ and $x_2,y_2,z_2$ represents the normalized state variables of the drive and driven systems, respectively. The terms $h(x_{1}), h(x_2)$ represents the three-segmented piecewise-linear function corresponding to the {\emph{Chua's diode}} as given in Eq. \ref{eqn:11}. The normalized parameters of the system are fixed as $\beta=1.152~, \nu=0.01388$, a=-1.0944, b=-0.5904, $\omega_{1,3}=2.1448,~ \omega_{2,4}=0.6627$. The ratio of the frequencies $\omega_{1,2}/\omega_{2,4}$ is an integral multiple of the {\emph{golden ratio}}. This system exhibits a prominent chaotic attractor obtained by the torus doubling or the {\emph{Heagy-Hammel}} route through the evolution of SNA at the values of the amplitude of the external forces $f_{1,3} = 0.5184,~f_{2,4} = 0.233$. The chaotic attractors of the drive (blue) and the driven (red) systems observed under the unsynchronized state $(\epsilon=0)$ is as shown in Fig. \ref{fig:12}.
\begin{figure}
\begin{center}
\includegraphics[scale=0.33]{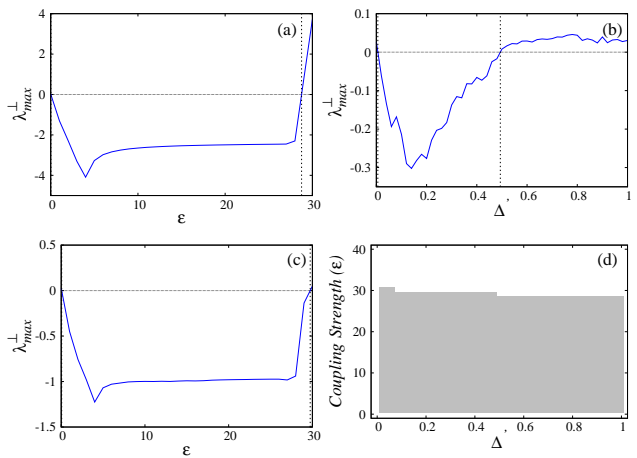}
\caption{(Color Online) Stability of synchronization observed in coupled quasiperiodically forced series LCR circuits with the {\emph{Chua's diode}} as the nonlinear element. (a) MSF obtained without transient uncoupling exists in negative valued regions in the range $0.0203 \le \epsilon \le 28.76$; (b) MSF as a function of $\Delta^{'}$ for $\epsilon = 29$ indicating stable synchronized states at the lower values of clipping fractions $0.006 \le \Delta^{'} \le 0.493$; (c) Variation of MSF with $\epsilon$ under transient uncoupling obtained with ${x_2}^{*} = 0.825$ and $\Delta^{'} = 0.4$ indicating enhanced stable synchronized states in the range of coupling strengths $0.056 \le \epsilon \le 29.738$; (d) Two-parameter bifurcation diagram in the $\Delta^{'}-\epsilon$ plane indicating the parameter regions for stable synchronized states (gray color).}
\label{fig:13}
\end{center}
\end{figure}
\begin{figure}
\begin{center}
\includegraphics[scale=0.66]{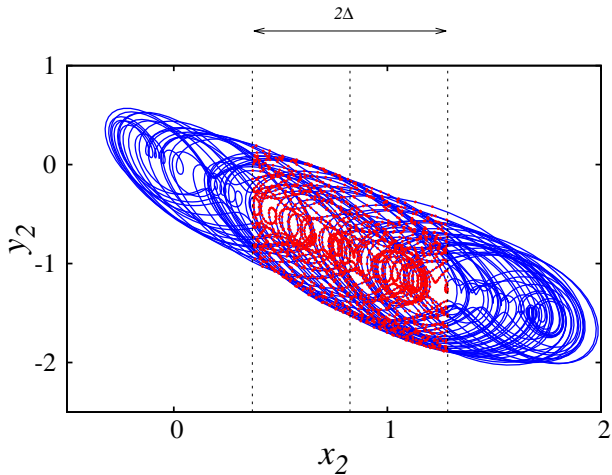}
\caption{(Color Online) Chaotic attractors of the driven quasiperiodically forced series LCR circuit system (blue) in the $x_2-y_2$ phase space with regions of $x_2$ (red) for which the real parts of the eigenvalues of the driven system is negative under a finite clipping fraction for the parameters ${x_2}^{*} = 0.825$ and $\Delta^{'} = 0.4$.}
\label{fig:14}
\end{center}
\end{figure}
The synchronization dynamics of the coupled system under transient uncoupling is shown in Fig. \ref{fig:13}. Under normal coupling, the variation of the MSF with the coupling strength indicates stable synchronized states in the range $0.0203 \le \epsilon \le 28.76$ as shown in Fig. \ref{fig:13}(a). The variation of the MSF of the coupled system under transient uncoupling as a function of the clipping fraction with the coupling strength fixed at $\epsilon = 29$ is shown in Fig. \ref{fig:13}(b). From Fig. \ref{fig:13}(b) it is observed that the coupled system exists in stable synchronized states in the lower regions of clipping fractions $0.006 \le \Delta^{'} \le 0.493$. Figure \ref{fig:13}(c) showing the variation of the MSF with the coupling strength indicates stable synchronized state in the range $0.056 \le \epsilon \le 29.738$ for a clipping fraction of $\Delta^{'}=0.4$. The two-parameter bifurcation diagram obtained in the $\Delta^{'}-\epsilon$ plane showing the parameter regions (gray) for the existence of the coupled system in the synchronized state is shown in Fig. \ref{fig:13}(d). \\
The Jacobian of the quasiperiodically forced series LCR circuit system is represented by the Eq. \ref{eqn:12}. The eigenvalues of the driven system is obtained using Eq. \ref{eqn:12} for the clipping fraction and coupling strength fixed at  $\Delta^{'}=0.4,~\epsilon=29$. Figure \ref{fig:14} shows the chaotic attractor (blue) of the driven system in the $(x_2-y_2)$ phase plane along with the values of $x_2$ (red) indicating the region of the negative eigenvalues. The eigenvalues corresponding to each of the piecewise-linear regions within the clipping fraction are either real or complex conjugates with negative real parts leading to the convergence of the trajectories of the driven system towards the drive.

\section{Conclusion}

In this paper we have reported the stability of induced synchronized states observed in unidirectionally coupled second-order chaotic systems subjected to transient uncoupling. The method of transient uncoupling has paved way for the enhancement of the stable synchronized states over higher values of coupling strength which are not observed under normal coupling. Further, a broader region of the clipping fraction over which stable synchronization is identified results in obtaining enhanced stability of synchronization over greater values of coupling strength. This method has been successfully applied for enhancing synchronization stability in higher dimensional chaotic systems \cite{Schroder2015,Schroder2016,Aditya2016,Ghosh2018}. The present study reveals the applicability of the method of transient uncoupling to second-order, non-autonomous systems exhibiting chaotic and strange non-chaotic behavior in their dynamics and establishes the global behavior of this method. All the simple chaotic systems present a broader region of clipping fraction over which stable synchronization is achieved. Further, an enhanced stable synchronization at higher values of coupling strength is obtained for a finite clipping fraction and the parameter regions of stable synchronization is identified. Because the number of components required to design the simple chaotic systems is minimum, these systems are much helpful in sustaining the synchronization stability over larger values of coupling strength for the practical applications of secure communication.

\bibliographystyle{apsrev4-1}
\bibliography{mybibfile}

\end{document}